\documentclass{article}
\usepackage{ltexpprt}

\usepackage{amsmath,amsfonts,graphpap,amscd,mathrsfs,graphicx,lscape}
\usepackage{epsfig,amssymb,amstext,xspace}
\usepackage{algorithm,algorithmic}

\usepackage{color}              
\usepackage{epsfig}

\newcommand{\E}{\mathbb{E}}
\newcommand{\abs}[1]{\vert{#1}\vert}

\def\squareforqed{\hbox{\rule{2.5mm}{2.5mm}}}

\def\QED{\ifmmode\squareforqed 
  \else{\nobreak\hfil   
    \penalty50                 
    \hskip1em                  
    \null                      
    \nobreak                   
    \hfil                      
    \squareforqed              
    \parfillskip=0pt           
    \finalhyphendemerits=0     
    \endgraf}                  
  \fi}

\def\blksquare{\rule{2mm}{2mm}}
\def\qedsymbol{\blksquare}
\newcommand{\bg}[1]{\medskip\noindent{\bf #1}}
\newcommand{\ed}{{\hfill\qedsymbol}\medskip}

\newenvironment{proofof}[1]{{\it{Proof of #1 : }}}{\ed}

\newcommand{\R}{\ensuremath{\mathbb R}}

\newcommand{\expect}[2]{\ensuremath{\text{{\bf E}$_{#1}\left[#2\right]$}}}

\newcommand{\comment}[1]{}
 \newcommand{\versions}[2]{#2}{}

\newcommand{\junk}[1]{}

\newcommand{\prob}{\mathbb{P}}

\newlength{\tmp} \newlength{\lpsx} \newlength{\lpsy} \newlength{\upsx} \newlength{\upsy}

\newcommand{\spe}{\text{\textsc{Spe}} }
\newcommand{\spoa}{\text{\textsc{PoA}} }

\newcommand{\poa}{\text{\textsc{PoA}} }

\newcommand{\opt}{\text{\textsc{Opt}} }
\newcommand{\vcg}{\text{\textsc{Vcg}} }

\newcommand{\matroid}{\ensuremath{\mathcal{M}=(\mathcal{E}_M,\mathcal{I}_M)}}

\begin{document}
\setcounter{page}{0}

\title{Sequential Auctions and Externalities}
%
%
%
%
%

%
\author{
Renato Paes Leme\thanks{ {\tt renatoppl@cs.cornell.edu} Dept of Computer
Science, Cornell University. Supported in part
by NSF grants CCF-0729006, and by internships at Google and Microsoft, and a Microsoft Fellowship.}
\and
Vasilis Syrgkanis \thanks{ {\tt vasilis@cs.cornell.edu} Dept of Computer
Science, Cornell University. Supported in part by ONR grant N00014-98-1-0589 and
NSF grants CCF-0729006, and by a Microsoft Internship.}
\and
\'{E}va Tardos \thanks{ {\tt eva@cs.cornell.edu} Dept of Computer
Science, Cornell University. Supported in part by NSF grants CCF-0910940 and
CCF-0729006, ONR grant N00014-08-1-0031,  a Yahoo!~Research Alliance Grant, and
a Google Research Grant.}
}
\date{}

\maketitle

\begin{abstract}

In many settings agents participate in multiple different auctions that are not
necessarily implemented simultaneously.  Future opportunities affect strategic
considerations of the players in each auction, introducing externalities.
Motivated by this consideration, we study a setting of a market of buyers and
sellers, where each seller holds one item, bidders have combinatorial valuations
and sellers hold item auctions sequentially.

Our results are qualitatively different from those of simultaneous auctions,
proving that simultaneity is a crucial aspect of previous work. We prove that if
sellers hold sequential  first price auctions then for unit-demand bidders
(matching market) every subgame perfect equilibrium achieves at least half of
the optimal social welfare, while for submodular bidders or when second price
auctions are used, the social welfare can be arbitrarily worse than the optimal.
We also show that a first price sequential auction for buying or selling a base
of a matroid is always efficient, and implements the VCG outcome.

An important tool in our analysis is studying first and second price auctions
with externalities (bidders have valuations for each possible winner outcome),
which can be of independent interest. We show that a Pure Nash Equilibrium
always exists in a first price auction with externalities.

%
%
\end{abstract}

\section{Introduction}

The first and second price auctions for a single item, and their corresponding strategically
equivalent ascending versions are some of the most commonly used auctions for
selling items. Their popularity has been mainly due to the fact that they
combine simplicity with efficiency: the auctions have
simple rules and in many settings lead to efficient allocation.

The simplicity and efficiency tradeoff is more difficult when auctioning
multiple items, especially so when items to be auctioned are possibly owned by
different sellers. The most well-known auction is the truthful VCG auction,
which is efficient, but is not simple: it requires coordination among sellers,
requires the sellers to agree on how to divide the revenue,
and in many situations requires solving computationally hard problems. In light
of these issues, it is important to design simple auctions with good
performance, it is also important to understand properties of simple auction designs used
in practice.

Several recent papers have studied properties of simple item-bidding auctions,
such as using simultaneous second price auctions for each item
\cite{Christodoulou2008,Bhawalkar}, or simultaneous first price auction
\cite{Hassidim11,Immorlica}. Each of these papers study the case when all the items are
auctioned simultaneously, a property essential for all of their results.
The simplest, most natural, and most common way to auction items by different owners
is to run individual single
item auctions (e.g., sell each item separately on eBay). No common auction environment is running simultaneous auctions (first price or second price) for large sets of items.
To evaluate to what
extent is the simultaneity important for the good properties of the above simple
auctions \cite{Christodoulou2008,Bhawalkar,Hassidim11}, it is important to
understand the sequential versions of item bidding auctions.

There is a large body of work on online auctions (see \cite{Parkes} for a
survey), where players have to make strategic decisions without having any
information about the future. In many auctions participants have information
about future events, and engage in strategic thinking about the upcoming
auctions. Here we take the opposite view from online auctions, and study the
full information version of this game, when the players have full information
about all upcoming auctions.


Driven by this motivation we study sequential simple auctions from an
efficiency perspective. Sequential auctions are very different from their
simultaneous counterparts. For example, it may not be dominated to bid above the
actual value of an item, as the outcome of this auction can have large effect
for the player in future auctions beyond the value of this item. We focus on two
different economic settings. In the first set of results we study the case of a
market of buyers and sellers, where each seller holds one item and each bidder
has a combinatorial valuation on the items. In the second setting we study the
case of procuring a base of a matroid from a set of bidders, each controlling an
element of the ground set.

For item auctions, we show that subgame perfect equilibrium always exists, and study the quality of the resulting outcome. While the equilibrium is not unique in most cases, in some important classes of games, the quality of any equilibrium is close to the quality of the optimal solution.
We show that in the widely studied case of matching markets
\cite{Shapley1971,Demange1986,Kranton2001}, i.e. when bidders are unit-demand,
the social welfare achieved by a subgame perfect equilibrium of a sequential
first price auction is at least half of the optimal.
Thus in a unit-demand setting
sequential implementation causes at most a factor of 2 loss in social welfare.
%
On the other hand, we also show that the welfare loss due to sequential implementation is unbounded
when bidders have arbitrary submodular valuations, or when the second price
auction is used, hence in these cases the simultaneity of the auctions is
essential for achieving the positive results
\cite{Christodoulou2008,Bhawalkar,Hassidim11}.

For the setting of auctioning a base of a matroid to bidders that control a
unique element of the ground set, we show that a natural sequential first price
auction has unique outcome, that achieves the same allocation and price outcome as $\vcg$.

An important building block in all of our results is a single item first or second price
auction in a setting with externalities, i.e., when players have different valuations for different
winner outcomes. Many economic settings might give rise to such
externalities among the bidders of an auction:
\begin{itemize}
\item We are motivated by externalities that arise in sequential auctions. In this setting, bidders might know of future auctions, and realize that their surplus in the future auction depends on who
wins the current auction. Such consideration introduces externalities, as players can have a different expected future utility according to the winner of the current auction.

To illustrate how externalities arise in sequential auction, consider the following example
of a sequential auction with two items and two players and with
valuations: $v_1(1) = v_1(2) = 5, v_1(1,2) = 10$ for player 1, and $v_2(1) = v_2(2) = v_2(1,2)
= 4$ for player 2. Player 1 has higher value for both item. However, if she wins item $1$ then she has
to pay 4 for the second item, while by allowing player 2 to win the first item,
results in a price of 0 for the second item. We can summarize this by saying
that his value for winning the first item is 6 (value 5 for the first item
itself, and an additional expected value of 1 for subsequently winning the
second item at a price 4), while her value for player 2 winning the first item is 5
(for subsequently winning the second item for free).
\item Bidders might want to signal information through their bids so as to
threat or inform other bidders and hence affect future options. This is the cause of the inefficiency in our example of sequential second price auction. Such phenomena
have been observed at Federal Communication Commission (FCC) auctions where
players were using the lower digits of their bids to send signals discouraging
other bidders to bid on a particular license (see chapter 1 of Milgrom
\cite{MilgromBook}).
\item If bidders are competitors or
collaborators in a market then it makes a difference whether your friend or
your enemy wins. One very vivid portrayal of such an externality is an auction
for nuclear weapons \cite{Jehiel96}.
\end{itemize}
The properties of such an auction are of independent interest outside of the
sequential auction scope.

\subsection{Our Results }

\quad

\noindent \textbf{Existence of equilibrium.} In section
\ref{sec:external-payoffs} we show that the first price single item auction
always has a pure Nash equilibrium that survives iterated elimination of
dominated strategies, even in the presence of
arbitrary externalities, strengthening the result in
\cite{JehielMoldovanuNonparticipation, Funk}.

\comment{First price auctions with externalities have been previously studies by
Jehiel and Moldovanu \cite{JehielMoldovanuNonparticipation} who showed that such
first price auctions have pure Nash equilibria, if bidders are allowed to use
dominated strategies. Funk \cite{Funk} proved that pure equilibria also exists
in undominated strategies. Our equilibrium construction also survives iterated
elimination of dominated strategies.}

In section \ref{sequential-item-auctions} we use such external auction to show
that sequential first price auctions have pure subgame perfect Nash equilibria
for any combinatorial valuations. This is in contrast to simultaneous first
price auctions that may not have pure equilibria even for very simple
valuations.

\textbf{Quality of outcomes in sequential first price item auctions.} Next we
study the quality of outcome of sequential first price auctions in various
settings. Our main result is that with unit demand bidders the price of
anarchy is bounded by 2. In contrast, when valuations are submodular, the price
of anarchy can be arbitrarily high. This differentiates sequential auctions from simultaneous
auctions, where pure Nash equilibria are socially optimal \cite{Hassidim11}.

These results extend to sequential auctions,
where multiple items are sold at each stage using independent first price auctions. Further, the efficiency
guarantee degrades smoothly as we move away from unit demand. Moreover, the results
also carry over to mixed strategies with a factor loss of $2$.
Unfortunately, the existence of pure equilibria is only guaranteed when
auctioning one item at-a-time.

\textbf{Sequential second price auctions.}
Our positive results depend crucially on using the first price auction format.
In the appendix, we show that sequential second price auctions can lead to
unbounded
inefficiency, even for additive valuations, while for additive valuations
sequential
first price, or simultaneous first or second price auctions all lead to
efficient outcomes.

\textbf{Sequential Auctions for selling basis of a matroid.} In section
\ref{sec:matroid} we consider matroid auctions, where bidders each
control a
unique element of the ground set, and we show that a natural sequential first
price
auction achieves the same allocation and price outcome as $\vcg$. Specifically,
motivated by the greedy spanning tree algorithm, we propose the following
sequential auction: At each iteration pick a cut of the matroid that doesn't
intersect previous winners and run a first price auction among the bidders in
the cut. This auction is a more distributed alternative to the ascending auction
proposed by Bikhchandani et al \cite{Bikhch2010}. For the interesting case of a
procurement auction for buying a spanning tree of a network from local
constructors, our mechanism takes the form of a
geographically local and simple auction.

We also study the case where bidders control several elements of the ground set
but have a unit-demand valuation. This problem is a common generalization of the matroid problem, and
the unit demand auction problem considered in the previous section. We show that the bound of 2 for the
price of anarchy of any subgame perfect equilibrium extends also to this generalization.

\subsection{Related Work}

\quad

\noindent \textbf{Externalities.} The fact that one player might influence the
other
without direct competition has been long studied in the economics literature.
The textbook model is due to Meade \cite{Meade} in 1952, and the concept has
been studied in various contexts. To name a few recent ones: \cite{Jehiel96}
study it in the context of weapon sales,
\cite{ghosh08,gomes09}  in the context of AdAuctions, and
\cite{Krysta10} in the context of combinatorial auctions. Our externalities
model is due to Jehiel and Moldovanu \cite{JehielMoldovanuNonparticipation}.
They show that a pure Nash
equilibrium exists in a full information game of first price auction with
externalities, but use dominated strategies in their proof. Funk \cite{Funk}
shows the existence of an equilibrium after one round of elimination of
dominated strategies, but argues that this refinement alone is not enough for
ruling out unnatural equilibria - and gives a very compelling example of this
fact. Iterative elimination of dominated strategies would
eliminate the unnatural equilibria in his example, but instead of analyzing it,
Funk analyzes a different concept: locally undominated strategies, which he
defines in the paper. We show the existence of an equilibrium surviving any
iterated elimination of dominated strategies - our proof is based on a
natural ascending price auction argument, that provides more intuition
on the structure of the game.
Our first price auction equilibrium is also an equilibrium in a second price auction. Previous work studying second price auctions include
Jehiel and  Moldovanu \cite{JehielMoldovanu00}, who study a simple case of
second price auctions with two buyers and externalities between the buyers,
derive equilibrium bidding strategies, and point out the various effects caused
by positive and negative externalities, while in \cite{JehielMoldovanu06} the
same authors study a simple case of second price auction with two types of
buyers.

\comment{
There is a long line of research in the economics
literature on auctions with externalities, see for example \cite{Greenwood91,
Jehiel96, Jehiel99}. It has also been studied in the context of AdAuctions
\cite{ghosh08,gomes09} and Combinatorial Auction \cite{Krysta10}. Variations of
our the single shot auctions with external payoffs appeared in \cite{Bae2009}
and \cite{Jehiel96}.
}

\textbf{Sequential Auctions.} A lot of the work in the economic
literature studies the behavior of prices in sequential auctions of identical
items where bidders participate in all auctions. Weber \cite{Weber2000} and
Milgrom and Weber \cite{Milgrom1982a} analyze first and second price sequential
auctions with unit-demand bidders in the Bayesian model of incomplete
information and show that in the unique symmetric Bayesian equilibrium the
prices have an upward drift. Their prediction was later refuted by
empirical evidence (see e.g. \cite{Ashenfelter1989}) that show a declining
price phenomenon. Several attempts to describe this ``declining price anomaly''
have since appeared such as McAfee and Vincent \cite{McAfee1993} that
attribute it to risk averse bidders. Although we study full information games
with pure strategy outcomes, we still observe declining price phenomena in
our sequential auction models without relying to risk-aversion.
Boutilier el al \cite{Boutilier99} studies first price auction in a setting with uncertainty, and gives a dynamic programming algorithm for finding optimal auction strategies assuming the distribution of other bids is stationary in each stage, and shows  experimentally that good quality solutions do emerge when all players use this algorithm repeatedly.

The multi-unit demands case has been studied under the complete
information model as well. Several papers (e.g. \cite{Gale2001,Rodriguez2009})
study the
case of two bidders. In the case of two bidders they show that there is a unique
subgame perfect equilibrium that survives the iterated elimination of weakly
dominated strategies, which is not the case for more than two bidders.  Bae et
al. \cite{Bae2009,Bae2008}
study the case of sequential second price auctions of identical items to two
bidders with concave valuations on homogeneous items. They show that the unique
outcome that
survives the iterated elimination of weakly dominated strategies is inefficient,
but achieves a social welfare at least $1-e^{-1}$ of the optimal. Here we
consider more than two bidders, which makes our analysis more challenging,
as the uniqueness argument of the Bae et al. \cite{Bae2009,Bae2008} papers
depends heavily on having only two players: when there are only two players, the
externalities that naturally arise due to the sequential nature of the auction
can be modeled by standard auction with no externalities using modified
valuations.

\textbf{Item Auctions.} Recent work from the
Algorithmic Game Theory community tries to propose the study of outcomes of
simple mechanisms for multi-item auctions.
Christodoulou, Kovacs and Schapira \cite{Christodoulou2008} and
Bhawalkar  and Roughgarden \cite{Bhawalkar} study the case of running
simultaneous second price item
auctions for combinatorial auction settings.  Christodoulou
et al. \cite{Christodoulou2008} prove that for bidders with submodular
valuations and incomplete information the Bayes-Nash Price of Anarchy is $2$.
Bhawalkar and Roughgarden \cite{Bhawalkar} study the more general case of
bidders with
subadditive valuations and show that under complete information the Price of
Anarchy of any Pure Nash Equilibrium is $2$ and under incomplete information the
Price of Anarchy of any Bayes-Nash Equilibrium is at most logarithimic in the
number of items.
Hassidim et al.  \cite{Hassidim11}  and Immorlica et al \cite{Immorlica} study the case of simultaneous first price auctions and show that the set of pure Nash equilibria of the game correspond to exactly the Walrasian equilibria. Hassidim et al.  \cite{Hassidim11}  also show that mixed Nash equilibria have a price of anarchy of 2 for submodular bidders and logarithmic, in the number of items, for subadditive valuations.

\textbf{Unit Demand Bidders.} Auctions with unit demand bidders correspond to the classical matching problem in optimization. They have been studied extensively also in the context of auctions, starting with the classical papers of Shapley and Shubik \cite{Shapley1971} and
Demange, Gale, and Sotomayor \cite{Demange1986}.
The most natural application of unit demand bidders is the case of a buyer-seller network market.
A different interesting application where sequential auction is also natural, is in the case of
scheduling jobs with deadlines. Suppose we have a set of jobs with different
start and end times (that are commonly
known) and each has a private valuation for getting the job done, not known to the auctioneer.
Running an auction for each time slot sequentially is natural since, for example, it
doesn't require for a job to participate in an auction before its start time.

\textbf{Matroid Auctions.} The most recent and related work on Matroid Auctions
is that of Bikhchandani et al \cite{Bikhch2010} who propose a centralized
ascending auction for selling bases of a Matroid that results in the \vcg
outcome. In their model each bidder has a valuation
for several elements of the matroid and the auctioneer is interested in selling
a base. Kranton and Minehart \cite{Kranton2001}
studied the case of a buyer-seller bipartite network market where each buyer has
a private valuation and unit-demand. They also propose an ascending auction that
simulates the \vcg outcome. Their setting can be viewed as a matroid auction,
where the matroid is the set of matchable bidders in the bipartite network.
Under this perspective their ascending auction is a special case of that of
Bikhchandani et al. \cite{Bikhch2010}. We study a sequential version of this
matroid basis auction game, but consider only the case when bidders are interested in a
specific element of the matroid, and show that the sequential auction also
implements the VCG outcome.

\section{Auctions with Externalities}\label{sec:external-payoffs}

In this section we consider a \emph{single-item auction with externalities}, and
analyze a simple first price auction for this case. Variations
on the same concept of externalities can be found, in Jehiel and Moldovanu
\cite{JehielMoldovanuNonparticipation}, Funk \cite{Funk} and in Bae et al
\cite{Bae2008} - the last one also motivated by sequential
auctions, but considered auctions with two players only. 
\comment{
Funk \cite{Funk} also gives an example of an equilibrium in undominated
strategies that leads to an unnatural outcome. In the example the unnatural
outcome would be eliminated by restricting to strategies that survive the
iterated elimination of dominated strategies. Instead, Funk introduces a new
concept of locally undominated strategy, and shows that first price auction with
externalities always has a pure Nash equilibrium using locally undominated
strategies.} 
Here we show that pure Nash equilibrium exists using strategies that survives
the iterated elimination of dominated strategies. 
The single-item auction with
externalities will be used as the main building block in the study of
sequential auctions.

\begin{Definition}
A \textbf{first-price single-item auction with externalities} with $n$
players is a game such that the type of player $i$ is a vector $[v_i^1,
v_i^2, \hdots, v_i^n]$ and the strategy set is $[0, \infty)$.  Given
bids $b=(b_1, \hdots, b_n)$ the first price auction allocates the item to the
highest bidder, breaking ties using some arbitrary fixed rule, and makes the
bidder pay his bid value. If player $i$ is allocated, then
experienced utilities are $u_i = v_i^i - b_i$ and $u_j = v_j^i$ for all $j \neq
i$. \end{Definition}

For technical reasons, we allow a player do bid $x$ and $x+$ for each real
number $x\geq 0$. The bid $b_i = x+$ means a bid that is infinitesimally larger
than $x$. This is essential for the existence of equilibrium in first-price
auctions. Alternatively, one could consider limits of $\epsilon$-Nash equilibria
(see Hassidim et al \cite{Hassidim11} for a discussion).
%
%


Next, we present a natural constructive proof of the existence of
equilibrium that survives iterated elimination of weakly-dominated strategies.
See the formal definition concept of iterated elimination of weakly-dominated
strategies we use in Appendix \ref{appendix:refinements}. Our proof is based on
ascending auctions.

\begin{theorem} \label{thm:external-exists}
 Each instance of the first-price single-item auction with externalities has
a pure Nash equilibrium that survives iterated elimination of weakly-dominated
strategies. \end{theorem}

\begin{proof}
For simplicity we assume that all $v_i^j$ are multiples of $\epsilon$. Further, we may
assume without loss of generality that $v\ge 0$ and $\min_j v_i^j=0$. We
say that an item is toxic, if $v_i^i < v_i^j$ for all $i \neq j$. If the item is
toxic, then $b_i = 0$ for all players and player $1$ getting the item is an
equilibrium. If not, assume $v_1^1 \geq v_1^2$.

 Let $\langle i,j,p \rangle$ denote the state of the game where player $i$ wins
for price $p$ and $j$ is the price setter, i.e., $b_i = p+$, $b_j = p$, $b_k =
0$ for $k \neq i,j$. The idea of the proof is to define a sequence of states
which have the following invariant property: $p \leq \gamma_i$, $p \leq
\gamma_j$ and $v_i^i -p \geq v_i^j$, where $\gamma_i = \max_j v_i^i - v_i^j$.
 \comment{We call the first two
conditions \emph{non-overbidding}.} We will define the sequence, such that
states don't appear twice on the sequence (so it can't cycle) and when the
sequence stops, we will have reached an equilibrium.

Start from the state $\langle 1,2,0 \rangle$, which clearly satisfies the
conditions. Now, if we are in state $\langle i,j,p \rangle$, if there is no $k$
such that $v_k^k - p > v_k^i$ then we are at an equilibrium satisfying all
conditions. If there is such $k$, move to the state $\langle k,i,p \rangle$ if
this state
hasn't appeared before in the sequence, and otherwise, move to $\langle
k,i,p+\epsilon \rangle$. We need to check that the new states satisfy the
invariant conditions: first $v_k^k - (p+\epsilon) \geq v_k^i$. Now, we need to
check the two first conditions: $p+\epsilon \leq v_k^k - v_k^i \leq \gamma_k$.
Now, the
fact that $i$ is not overbidding is trivial for $\langle k,i,p \rangle$, since
$i$ wasn't overbidding in $\langle i,j,p \rangle$. If, $\langle k,i,p \rangle$
already appeared in this sequence, it means that $i$ took the item from player
$j$, so: $v_i^i - p > v_i^j$ so: $p < v_i^i - v_i^j \leq \gamma_i$ so
$p+\epsilon \leq \gamma_i$.

Now, notice this sequence can't cycle, and prices are bounded by valuations, so
it must stop somewhere and there we have an equilibrium. To show the existence
of an equilibrium surviving iterative elmination of weakly dominated
strategies, we need a more careful argument: we refer to appendix
\ref{appendix:refinements} for a proof.
\end{proof}

It is not hard to see that any equilibrium of the \emph{first-price} auction
with externalities is also an equilibrium in the \emph{second-price} version.
The subclass of second-price equilibria that are equivalent to a
first-price equilibria (producing same price and same allocation),
are the second-price equilibria that are \textbf{envy-free}, i.e., no player
would rather be the winner by the price the winner is actually paying. So, an
alternative way of looking at our results for first price is to see them as
outcomes of second-price when we believe that envy-free equilibria are selected
at each stage.

\section{Sequential Item Auctions}\label{sequential-item-auctions}

Assume there are $n$ players and $m$ items and each
player has a monotone (free-disposal)  combinatorial valuation $v_i: 2^{[m]}
\rightarrow \R_+$. We will consider sequential auctions. First assume that at each time step only a single item is being auctioned off: item $t$ is auctioned in step $t$.
We define the sequential first (second) price auction for this case as
follows: in time step $t = 1 \hdots m$ we ask for bids $b_i(t)$ from each agent for the item being considered in this step,
and run a first (second) price auction to sell this item. Generally, we assume that after
each round, the bids of each agent become common knowledge, or at least the winner and the winning price become public knowledge. The agents can
naturally choose their bid in time $t$ as a function of the past history.
We will also consider the natural extension of these games, when each round can
have multiple items on sale, bidders submit bids for each item on sale, and we
run a separate first (second) price auction for each item.

This setting is captured by \textbf{extensive form games} (see \versions{the
Appendix of the full version}{Appendix \ref{appendix:extensive} } for a formal
definition and 
\cite{fudenberg91} for a more comprehensive treatment). The strategy of each
player is an adaptive bidding policy: the policy specifies what a player bids
when the $t^{th}$ item (or items) is auctioned, depending on the bids and outcomes
of the previous $t-1$ items. More formally a strategy for player $i$ is a bidding function
$\beta_i(\cdot)$ that associates  a bid $\beta_i(\{b_i^\tau\}_{i,\tau <
t})\in \R_+$ with each sequence of previous bidding
profiles $\{b_i^\tau\}_{i,\tau < t}$.

Utilities are calculated in the natural way: utility for the set of items won,
minus the sum of the payments from each round. In each round the player with
largest bid wins the item and pays the first (second) price. We are interested
in the \textbf{subgame perfect equilibria} ($\spe$) of this game: which means
that the
profile of bidding policies is a Nash equilibrium of the original
game and if we arbitrarily fix what happens in the first $t$ rounds, the policy
profile also remains a Nash equilibrium of this induced game.

Our goal is to measure the \textbf{Price of Anarchy}, which is the worse
possible ratio between the optimal welfare achievable (by allocating the items
optimally) and the welfare in a subgame perfect equilibrium. Again, we invite
the reader to \versions {the Appendix of the full version}{Appendix \ref{appendix:extensive}} for formal definitions.


\subsection{First Price Auctions: existence of pure equilibria}\label{sec:first-price}

First we show that sequential first price single item auctions have pure equilibria for all valuations.

\begin{theorem}
Sequential first price auction when each round a single item is auctioned has a $\spe$ that doesn't use dominated strategies, and in which bids in each node of the game tree depend only on who got the item in the previous rounds.
\end{theorem}

We use backwards induction, and apply our result on the existence of Pure Nash
Equilibria in first price auctions with externalities to show the theorem.
Given outcomes of the game starting from stage $k+1$ define a game with externalities for stage $k$, and by Theorem \ref{thm:external-exists} this game has a pure Nash equilibrium. It interesting to
notice that we have existence of a pure
equilibrium for arbitrary combinatorial valuation. In contrast, the
simultaneous item bidding auctions, don't always possess a pure equilibrium even for subadditive bidders (\cite{Hassidim11}).

\comment{
We show that in sharp contrast to second price, sequential first price auctions
have a small price of anarchy. The main reason is that equilibria are envy-free,
i.e. if some player is winning some item for a certain price $p$, other player
is able to take this item for the same price $p$ that he is currently paying.

Consider for example players with additive valuations and let's look at the
non-overbidding equilibria, in the sense that players don't overbid in the
``auction with external payoffs'' induced in each node -- notice that it is
still possible that a player bids more than his marginal value for some item. In
this setting, there only one thing that can happen for the last item, it is sold
for the player that values it the most by the second highest price (that is the
only first-price equilibrium where players don't overbid). Now, since the last
auction is determined regardless of what happens before, there is one possible
equilibrium for the before to last item, which is again the player that values
it the most wins for the second highest price and so on... Therefore, the only
subgame perfect equilibrium is efficient.
}

In the remainder of this section we consider three classes of valuations:
additive, unit-demand and submodular. For additive valuations, the sequential
first-price auction always produces the optimal outcome. This is in contrast to
second price auctions, as we show in \versions{the Appendix of the full version}{Appendix  \ref{sec:second-price}}.

In the next two subsections we consider unit-demand bidders, and prove a bound of 2 for the Price of Anarchy, and then show that for submodular valuations the price of anarchy is unbounded (while in the simultaneous case, the price of anarchy is bounded by 2 \cite{Hassidim11}).

\comment{
We present our results for the case of one item being auctioned at each
timestep. However, all results carry over to the case where in each timestep a
subset of items is auctioned simultaneously -- i.e. an intermediate model
between our fully sequential model and the fully simultaneous model of
\cite{Hassidim11}. However, if multiple items are auctioned in a round, a
subgame perfect equilibrium might fail to exist, as shown in \versions{Appendix of the full version}{ appendix
\ref{appendix:multi-unit}}.

Consider players with additive valuations and let's look at the
non-overbidding equilibria, in the sense that players don't overbid in the
``auction with external payoffs'' induced in each node -- notice that it is
still possible that a player bids more than his marginal value for some item. In
this setting, there only one thing that can happen for the last item, it is sold
for the player that values it the most by the second highest price (that is the
only first-price equilibrium where players don't overbid). Now, since the last
auction is determined regardless of what happens before, there is one possible
equilibrium for the before to last item, which is again the player that values
it the most wins for the second highest price and so on... Therefore, the only
subgame perfect equilibrium is efficient. In section \ref{sec:second-price} we
show that this is not true for second-price.
}

\subsection{First Price Auction for Unit-Demand Bidders}\label{subsec:multi_unit_auctions}
We assume that there is free disposal, and hence say that a player $i$ is unit-demand if, for a bundle $S \subseteq [m]$, $v_i(S) = \max_{j \in S} v_{ij}$, where $v_{ij}$ is the valuation of player $i$ for item $j$.


To see that inefficient allocations are possible, consider the example given in
Figure \ref{fig1}. There is a sequential first price auction of three items
among four players. Player $b$ prefers to loose the first item, anticipating
that he might get a similar item for a cheaper price later. This gives an example
where the Price of Anarchy is $3/2$. Notice that this is the only equilibrium using
non-dominated strategies.

\versions{
\begin{figure}
\centering
\includegraphics{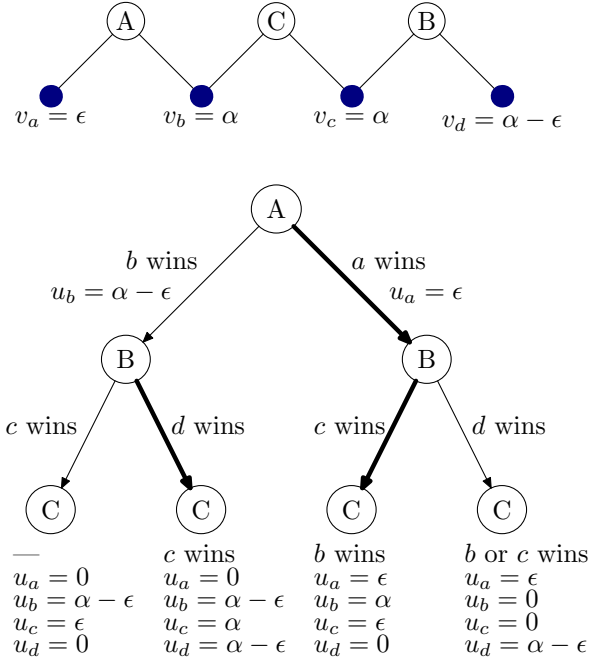}
\caption{Sequential Multi-unit Auction generating $\poa$ $3/2$: there
are $4$ players $\{a,b,c,d\}$ and three items that are auctioned first $A$, then
$B$ and then $C$. The optimal allocation is $b\rightarrow A$, $c\rightarrow C$,
$d \rightarrow B$ with value $3\alpha-\epsilon$. There is a $\spe$ that has
value $2\alpha+\epsilon$. In the limit when $\epsilon$ goes to $0$ we get
$\poa=3/2$.
}
\label{fig1}
\end{figure}
}{
\begin{figure}[h]
\centering
\includegraphics{unit-demand-example1.mps}
\caption{Sequential Multi-unit Auction generating $\poa$ $3/2$: there
are $4$ players $\{a,b,c,d\}$ and three items that are auctioned first $A$, then
$B$ and then $C$. The optimal allocation is $b\rightarrow A$, $c\rightarrow C$,
$d \rightarrow B$ with value $3\alpha-\epsilon$. There is a $\spe$ that has
value $2\alpha+\epsilon$. In the limit when $\epsilon$ goes to $0$ we get
$\poa=3/2$.
}
\label{fig1}
\end{figure}
}

\begin{theorem}\label{thm:unit-demand}
For unit-demand bidders, the $\poa$ of pure subgame perfect equilibria of Sequential First Price Auctions of individual items is bounded by $2$, while for mixed equilibria it is at most $4$.
\end{theorem}

\begin{proof}
 Consider the optimal allocation and a subgame perfect equilibrium, and let
$\opt$ denote the social value of the optimum, and $\spe$ the social value of the subgame perfect equilibrium. Let $N$ be the set of players allocated in the optimum.
For each $i\in N$, let $j^*(i)$ be the element it was allocated to in the
 optimal, and let $j(i)$ be the element he was allocated in the
subgame perfect equilibrium  and let
 $v_{i,j(i)}$ be player $i$'s value for this element (if player $i$ got more
than one element, let $j(i)$ be his most valuable element). If player $i$ wasn't
allocated at all, let $v_{i,j(i)}$ be zero. Let $p(j(i))$ be the price for which
item $j(i)$ was sold in equilibrium. Consider three possibilities:

\begin{enumerate}
 \item $i$ gets $j^*(i)$, then clearly $v_{i,j(i)} \geq v_{i,j^*(i)}$
 \item $i$ gets $j(i)$ after $j^*(i)$ or doesn't get allocated at all, then
  $v_{i,j(i)} \geq v_{i,j^*(i)} - p(j^*(i))$, otherwise
  he could have improved his utility by winning $j^*(i)$
 \item $i$ gets $j(i)$  before $j^*(i)$, then either $v_{i,j(i)} \geq
v_{i,j^*(i)}$ or he can't improve his utility by getting $j^*(i)$, so it
must be the case that his marginal gain from $j^*(i)$ was smaller than
the maximum bid in $j^*(i)$, i.e. $p(j^*(i)) \geq v_{i,j^*(i)} - v_{i,j(i)}$
\end{enumerate}
Therefore, in all the cases, we got $p(j^*(i)) \geq v_{i,j^*(i)} - v_{i,j(i)}$.
Summing for all players $i\in N$, we get:
$$\opt = \sum_i v_{i,j^*(i)} \leq \sum_i v_{i,j(i)} + p(j^*(i)) \leq 2
\spe$$
where in the last inequality is due to individual rationality of the players.

\versions{The bound of 4 for mixed equilibria is proved in the full version.
}
{Next we prove the bound of 4 for the mixed case.
We focus of a player $i$ and let $j=j^*(i)$ denote item assigned to $i$ in the optimal matching.
In the case of mixed Nash equilibria, the price $p(j)$ is a random variable, as well as $A_i$ the set of items  player $i$ wins in the auction. Consider a node $n$ of the extensive form game, where $j$ is up for auction, i.e., a possible history of play up to $j$ being auctioned. Let $P^{n-}_i$ be the expected value of the total price $i$ paid till this point in the game, and let
$P^n(j)=\expect{}{p(j)|n}$ be the expected price for item $j$ at this node $n$, and note that $P(j)=\expect{}{p(j)}=\expect{}{P^n(j)}$, where the right expectation is over the induced distribution on nodes $n$ where $j$ is being auctioned.

Player $i$ deviating by offering price $2 P^n(j)$ at every node $n$ that $j$ comes up for auction, and then dropping out of the auction, gets him utility at least $1/2(v_i(j)-2P^n(j))-P^{n-}_i$, as he wins item $j$ with probability at least 1/2 and paid $P^{n-}_i$ to this point. Using the Nash inequality we get
$$\expect{}{v_i(A_i)}-P_i \ge \expect{}{1/2(v_i(j)-2P^n(j))-P^{n-}_i},$$
where $P_i$ is the expected payment of player $i$, and the expectation on the right hand side is over the induced distribution on the the nodes of the game tree where $j$ is being auctioned.
Note that the proposed deviation does not effect the play before item $j$ is being auctioned, so the expected value of $\expect{}{P^{n-}_i}$ over the nodes $n$ is at least the expected payment $P_i$ of player $i$, and that the expected value of $P^n_j$ over the nodes $u$ is the expected price $P(j)$ of item $j$. Using these we get
$$\expect{}{v_i(A_i)}\ge \frac{1}{2}v_i(j)-P(j).$$
Now summing over all players, and using that $\sum_j P(j) \le \sum_i \expect{}{v_i(A_i)}$ due to individual rationality, we get the claimed bound of 4.}

\end{proof}

The proof naturally extends to sequential auctions when in each round multiple items are being auctioned.
%
%
We can also generalize the above positive result to any
class of valuation functions that satisfy the property that the optimal
matching allocation is close to the optimal allocation.

\begin{theorem}
\label{thm:approx-unit}
Let $\opt_M$ be the optimal matching allocation and $\opt$ the optimal
allocation of a Sequential First Price Auction. If $\opt \leq \gamma \opt_M$
then the $\spoa$ is at most $2\gamma$ for pure equilibria and at most $4\gamma$ for mixed Nash, even
 if each round multiple items are auctioned in parallel (using separate first price auctions).
\end{theorem}

\versions{}{
\begin{proof}
 Let $j^*(i)$ be the item of bidder $i$ in the optimal matching allocation and
$A_i$ his allocated set of items in the $\spe$. Let $A_i^-$ be the items that
bidder $i$ wins prior or concurrent to the auction of $j^*(i)$ and $A_i^+$ the ones that he
wins after. Consider a bidder $i$ that has not won his item in the optimal matching allocation. Bidder $i$
could have won this item when it appeared by bidding above its current
price $p_{j^*(i)}$ and then abandon all subsequent auctions. Hence:

 $$ v_i(A_i^- \cup \{j^*(i)\})- p_{j^*(i)}-\sum_{j\in A_i^-}p_j \leq
v_i(A)-\sum_{j\in A}p_j $$
$$ v_i(A_i^- \cup \{j^*(i)\})-p_{j^*(i)} \leq v_i(A)-\sum_{j\in A_i^+}p_j \leq
v_i(A) $$
$$v_i(j^*(i))-p_{j^*(i)} \leq v_i(A)$$

If a player did acquire his item in the optimal matching allocation then the above inequality certainly
holds. Hence, summing up over all players we get:
\begin{equation*}
\begin{split}
 \opt_M = & \sum_i v_i(j^*(i)) \leq \sum_i v_i(A_i) +\sum_i
p_{j^*(i)} \\
\leq & \spe + \sum_{j} p_j = \spe + \sum_i \sum_{j\in A_i}p_j \\
\leq & \spe + \sum_i v_i(A_i) = 2\spe
\end{split}
\end{equation*}
which in turn implies: $$\opt\leq \gamma \opt_M\leq 2\gamma \spe$$

The bound of $4\gamma$ for the mixed case is proved along the lines as the mixed proof of
Theorem \ref{thm:unit-demand}.
\end{proof}

The above general result can be applied to several natural classes of bidder
valuations. For example, we can derive the following corollary for
multi-unit auctions with submodular bidders: a bidder is said
to be uniformly submodular if his valuation is a submodular function on the
number of items he has acquired and not on the exact set of items. Thus a
submodular valuation is defined by a set of decreasing marginals
$v_i^1,\ldots,v_i^m$.

\begin{corollary}
If bidders have uniformly submodular valuations and $\forall i,j:
|v_i^1-v_j^1|\leq \delta \max(v_i^1,v_j^1)$ ($\delta<1$) and there are more
bidders than items then the $\spoa$ of a Sequential First Price Auction is at
most $2/(1-\delta)$.
\end{corollary}

}


\subsection{First Price Auctions, Submodular Bidders}\label{subsec:submodular}

In sharp contrast to the simultaneous item bidding auction, where both first
and second price have good price of anarchy whenever Pure Nash equilibrium exist
\cite{Christodoulou2008, Bhawalkar, Hassidim11}, we show that for certain
submodular valuations, no welfare guarantee is possible in the sequential case.
While there are multiple equilibria in such auctions, in our example the natural
equilibrium is arbitrarily worse then the optimal 
allocation.

\begin{theorem}\label{submodular_thm}
For submodular players, the Price of Anarchy of the sequential first-price
auction is unbounded.
\end{theorem}

\versions{
The example with unbounded Price of Anarchy are discussed in the full version.}{}
The intuition is that there is a misalignment between social welfare
and player's utility. A player might not want an item for which he has high
value but has to pay a high price. In the sequential setting, a bidder may
prefer to let a
smaller value player win because of the benefits she can derive from his decreased value on future items, allowing her to buy future items at a smaller price, or diverting a competitor, and hence decreasing the price.

\versions{}{
\begin{proof}
Consider four players and $k+3$ items where $2$
of the payers have additive valuations and $2$ of them has a coverage function
as a valuation. Call the items $\{I_1, \hdots, I_k, Y, Z_1, Z_2\}$ and let
players $1,2$ have additive valuations. Their valuations are represented by the
following table:

\begin{center}
\begin{tabular}{ c || c | c | c | c | c | c  }
    & $I_1$ & $\hdots$ & $I_k$ & $Y$ & $Z_1$ & $Z_2$ \\
  \hline
  $1$ & $1+\epsilon$ & $\hdots$ & $1+\epsilon$ & $0$ & $2-k\delta/2$ &
$0$ \\
  $2$ & $1$ & $\hdots$ & $1$ & $0$ & $0$ & $2-k\delta/2$ \\
\end{tabular}
\end{center}
The valuations of players $3$ and $4$ are given by the coverage
functions defined in Figure \ref{fig_vals}: each item corresponds to a set
in the picture. If the player gets a set of items, his valuation for those items
(sets)
is the sum of the values of the elements covered by the sets corresponding to
the items.

\begin{figure}
\centering
\includegraphics{coverage_v41.mps}
\caption{Valuations $v_3$ and $v_4$.}
\label{fig_vals}
\end{figure}

In the optimal allocation, player $1$ gets all the items $I_1, \hdots, I_k$,
player $3$ gets $Y$ and player $4$ gets $Z_1, Z_2$. The resulting social welfare
is $k + 8 + k\epsilon - \delta/2$. We will show that there is a subgame perfect
equilibrium such that player $3$ wins all the items $I_1, \hdots, I_k$, even
though it has little value for them, resulting in a social welfare of
approximately 8 only.
The intuition is the following: in the end
of the auction, player $4$ has to decide if he goes for item $Y$ or goes for
items
$Y_1, Y_2$. If he goes for item $Y$, he competes with player $3$ and afterwards
lets players $1$ and $2$ win items $Z_1, Z_2$ for free. This decision of player
$4$
depends on the outcomes of the first $k$ auctions. In
particular, we show that if all items $I_1, \hdots, I_k$ go to either $3$ or
$4$, then player $4$ will go for item $Y$, otherwise, he will go for items $Z_1,
Z_2$. If either players $1$ or $2$ acquire any of the items  $I_1, \hdots,
I_k$,
they will be guaranteed to lose
item $Z_1, Z_2$, and therefore both will start bidding truthfully on all
subsequent $I_i$ auctions, deriving very little utility. In equilibrium
agent $3$ gets all items $I_1, \hdots, I_k$, resulting in a social welfare of
approximately $8$ only.

In the remainder of this section, we provide a more formal analysis:  We
begin by examining what happens in the last three auctions of $Y,Z_1$ and $Z_2$
according to what happened in the first $k$ auctions. Let $k_{1,2},k_3,k_4$ be
the number of items won by the corresponding players in the first $k$ auctions.
\begin{itemize}
 \item Case 1: $k_{1,2}=0$. Thus $k_3 = k-k_4$. Player $3$ has a value of
$4-\frac{\delta}{2}-k\delta+(k-k_3)\delta = 4-\frac{\delta}{2}-(k-k_4)\delta$
for item $Y$. Player $4$ has a value of $4$ for $Y$ and a value of
$2-k\frac{\delta}{2}+(k-k_4)\frac{\delta}{2}$ for each of $Z_1$ and $Z_2$. Thus
if player $4$ loses auction $Y$ he will get a utility of $(k-k_4)\delta$ from
the auctions of $Z_1$ and $Z_2$ since players $1$ and $2$ will bid
$2-k\delta/2$. Thus at auction $Y$ player $4$ is willing to win for a price of
at most $4-(k-k_4)\delta$ and player $3$ will bid
$4-\frac{\delta}{2}-(k-k_4)\delta$.
Thus, player $4$ will win $Y$ and will only bid $(k-k_4)\frac{\delta}{2}$ in
each of $Z_1,Z_2$. Therefore, in this case we get that the utilities of all the
players from the last three auctions are:
$u_1=u_2=2-(2k-k_4)\frac{\delta}{2},u_3=0,u_4=\frac{\delta}{2}+(k-k_4)\delta$
 \item Case 2: $k_{1,2}>0$. Player $3$ has a value of
$4-\frac{\delta}{2}-k\delta+(k-k_3)\delta$ for $Y$. Since $k_{12}\geq 1$, we
have $k-k_3=k_4+k_{1,2}\geq k_4+1$, hence the value of $3$ for $Y$ is at least
$4+\frac{\delta}{2}-(k-k_4)\delta$. Player $4$ has a value of $4$ for $Y$
and a value of $2-k\frac{\delta}{2}+(k-k_4)\frac{\delta}{2}$ for each of $Z_1$
and $Z_2$. Hence, again player $4$ wants to win at auction $Y$ for at most
$4-\frac{\delta}{2}-(k-k_4)\delta$, hence he will lose to $3$ and will go on to
win both $Z_1$ and $Z_2$.
Thus the utilities of all the
players from the last three auctions are:
$u_1=u_2=0,u_3=\delta,u_4=(k-k_4)\delta$
\end{itemize}

We show by induction on $i$ that as long as players $1$ and $2$ haven't won any
of the $k-i$ items auctioned so far then they will bid $0$ in the remaining $i$
items and one of players $3$ or $4$ will win marginally with zero profit.
For $i=1$ since both $1$ and $2$ haven't won any previous item, by losing the
$k$'th item we know by the above analysis that they both get utility of
$\approx 2$, while if any of them wins then they get utility of $0$.
The external auction that is played at the $k$'th item is represented by the
following $[v^i_j]$ matrix:
$$ \begin{bmatrix} 1+\epsilon & 0 & \approx 2 & \approx 2\\
                 0 & 1 & \approx 2 & \approx 2\\
		 \delta & \delta & \delta & 0 \\
		 (k-k_4)\delta & (k-k_4)\delta &
\frac{\delta}{2}+(k-k_4)\delta &
\frac{3\delta}{2}+(k-k_4)\delta \end{bmatrix} $$
It is easy to observe that the following bidding profile is an equilibrium of
the above game that doesn't involve any weakly dominated strategies:
$b_1=b_2=0,b_3=\delta,b_4=\delta+$. Thus, player $4$ will marginally win with
no profit from the current auction (alternatively we could have player $3$ win
with no profit).

Now we prove the induction step. Assume that it is true for the $i-1$. We know
that if either player $1$ or $2$ wins the $k-i$ item then whatever they do in
subsequent auctions, from the case 2 of the analysis, player $4$ will go for
$Z_1$ and $Z_2$ and they will get $0$ utility in the last $3$ auctions. Hence,
in the $i-1$ subsequent auctions they will bid truthfully, making player $1$
win marginally at zero profit every auction. On the other hand if they lose, by
the induction hypothesis they will lose all subsequent auctions leading them to
utility of $\approx 2$. Moreover, players $3$ and $4$ have the same exactly
utilities as in the base case, since they never acquire any utility from the
first $k$ auctions. Thus the external auction played at the $k-i$ item is
exactly the same as the auction of the base case and hence has the same bidding
equilibrium.

Thus in the above $\spe$ players $1$ and $2$ let some of the players $3$ and
$4$ win all the first $k$ items. This leads to an unbounded $\spoa$.
\end{proof}
}

\section{Matroid Auctions}
\label{sec:matroid}

In this section we first consider a matroid auction where each matroid element
is associated with a separate bidder, then in Section
\ref{subsec:sequential_matroid} we consider a problem that generalizes matroid
auctions and item auctions with unit demand bidders.

\subsection{Sequential Matroid Auctions}
\label{subsec:sequential_matroid}

Suppose that a telecommunications company wants to build a
spanning tree over a set of nodes. At each of the possible links of the network
there is a distinct set of local constructor's that can build the link. Each
constructor has a private cost for building the link. So the company has to
hold a procurement auction to get contract for building edges of a spanning tree with
minimum cost.
In this section, we show that by running a sequential first price auction,
we get the outcome equivalent to the $\vcg$ auction in a distributed and
asynchronous fashion.

A version of the well-known greedy algorithm for this optimization problem is to
consider cuts of this graph sequentially, and for each cut we consider, include
the minimum cost edge of the cut.
Our sequential auction is motivated by this greedy algorithm:
we run a sequence of first price auctions among the edges in a cut. More
formally, at each stage of the auction, we consider a cut where no edge was
included so far, and hold a first price sealed bid auction, whose winner is
contracted.
More generally, we can run the same auction on any matroid, not just the
graphical matroid considered above. The goal of the procurement auction is to
select a minimum cost matroid basis, and at each stage we run a sealed bid first
price auction for selecting an element in a co-circuit.

Alternately, we can also consider the analogous auction for selling some service to a basis of a matroid. As before, the bidders correspond to elements of a matroid. Their private value $v_i$ is their value for the service. Due to some conflicts, not all subsets of the bidders can be selected. We assume that feasible subsets form a matroid, and hence the efficient selection chooses the basis of maximum value. As before, it may be simpler to implement smaller regional auctions. Our method sequentially runs first price auctions for adding a bidder from a
co-circuit.  For the special case of the dual of graphical matroid, this problem corresponds to the following.  Suppose that a telecommunications company due to some mergers ended up with
a network that has cycles. Thus the company decides to
sell off its redundant edges
so that it ends up owning just a spanning tree. The sequential auction  we propose
runs a sequence of first price auctions, each time selecting an edge of a cycle in the network for sale. If more than one bidder is interested in an
edge we can simply think of it as replacing that edge with a path of edges, each
controlled by a single individual.

The main result of this section is that the above sequential auction implements
the $\vcg$ outcome both for procurement and direct auctions.
To unify the presentation with the other sections, we will focus here on direct
auction. In the final subsection, we will consider a common generalization of
the unit-demand auction and this matroid auction. In the procurement version, we
make the small technical assumption that every cut of the matroid contains at
least two elements, otherwise the $\vcg$ price of a player could be
infinity. Such assumption
was also made in previous work on matroid auctions \cite{Bikhch2010}.

\begin{theorem} In a sequential first price auction among players in the
co-circuit of a matroid (as described above), subgame perfect equilibria in
undominated strategies emulate the $\vcg$ outcome (same allocations and prices).
\label{thm:matroid_spe_opt}
\end{theorem}

\comment{
For completeness, we summarize definitions regarding matroids in \versions{the Appendix of the full version}{Appendix \ref{apdx:matroid}}. Here we summarize the notation we need.
Let $\mathcal M$ denote a matroid on ground set $\mathcal{E}_M$, we use $\mathcal{I}_M$ to denote set of independent sets of $\mathcal M$, and $\mathcal{C}(\mathcal{M})$ the set of circuits of $\mathcal M$. For
a subset $S\subset \mathcal{E}_M$ we use $r_{\mathcal{M}}(S)$ to denote the
rank of the set $S$, and use  $\mathcal{M}/S$ to denote the matroid with set $S$ contracted.
}

For completeness, we summarize some some definitions regarding matroids and
review notation.

A Matroid $\mathcal{M}$ is a pair
$(\mathcal{E}_M,\mathcal{I}_M)$, where $\mathcal{E}_M$ is a ground set of
elements and $\mathcal{I}_M$ is a set of subsets of $\mathcal{E}_M$ with the
following properties: (1) $\emptyset \in \mathcal{I}_M$, (2) If $A\in
\mathcal{I}_M$ and $B\subset A$ then $B\in \mathcal{I}_M$, (3) If $A,B \in
\mathcal{I}_M$ and $|A|>|B|$ then $\exists e\in A-B$ such that $B+e \in
\mathcal{I}_M$. The subsets of $\mathcal{I}_M$ are the independent subsets of
the matroid and the rest are called dependent.

The rank of a set $S\subset \mathcal{E}_M$, denoted as
$r_{\mathcal{M}}(S)$, is the cardinality of the maximum independent subset of
$S$. A base of a matroid is a maximum cardinality independent set and the set of
bases of a matroid $\mathcal{M}$ is denoted with $\mathcal{B}_{\mathcal{M}}$.
the
An important example of a matroid is the graphical matroid on the edges of a
graph, where a set $S$ of edges is independent if it doesn't contain a cycle,
and bases of this matroid are the spanning trees.

A circuit of a matroid $\mathcal{M}$ is a minimal dependent set.
and we denote with $\mathcal{C}(\mathcal{M})$ the set of circuits of
$\mathcal{M}$.
Circuits in graphical matroids are exactly the cycles of the graph. A cocircuit
is a minimal set that intersects every base of $\mathcal{M}$. Cocircuits in
graphical matroids corresond to cuts of the graph.

\begin{Definition}[Contraction]
 Given a matroid $\matroid$ and a set $X\subset \mathcal{E}_M$ the contraction
of $\mathcal{M}$ by $X$, denoted $\mathcal{M}/X$, is the matroid defined on
ground set $\mathcal{E}_M-X$ with $\mathcal{I}_{M/X}=\{S\subseteq
\mathcal{E}_M-X: S\cup X \in \mathcal{I}_M\}$.
\end{Definition}

If we are given weights for each element of the ground set of a matroid
$\mathcal{M}$ then it is natural to define the following optimization problem:
Find the base $\opt(\mathcal{M})\in \mathcal{B}_{\mathcal{M}}$ that
has minimum/maximum total weight (we might sometimes abuse notation and denote
with $\opt$ both the set and its total weight). A well known algorithm for
solving the above
optimization problem is the following (see \cite{Lawler1976}): At each iteration
consider a cocircuit that doesn't intersect the elements already picked in
previous iterations, then add its minimum/maximum element to the current
solution.


The most well known mechanism for auctioning items to a set of bidders is
the Vickrey-Clarke-Groves Mechanism ($\vcg$). The
\vcg mechanism selects the optimal basis $\opt(\mathcal{M})$. It is easy to see that
the \vcg price of a player $i\in \opt(\mathcal{M})$, denoted as $\vcg_i(\mathcal{M})$, is
the valuation of the highest bidder $j(i)$ that can be exchanged with $i$ in $\opt(\mathcal{M})$,
i.e. $\vcg_i(\mathcal{M})=\max\{v_j: \opt(\mathcal{M})-i+j\in \mathcal{I}_M\}$, or alternately
the above price is the maximum over all cycles of the matroid that contain $i$
of the minimum value bidder in each cycle: $\vcg_i(\mathcal{M}) = \max_{C\in
\mathcal{C}(\mathcal{M}): i\in C}\min_{i\neq j\in C} v_j$. To unify notation we
say that $\vcg_i(\mathcal{M})=\infty$ for a bidder $i\notin
\opt(\mathcal{M})$, although the actual price assigned by the \vcg mechanism is
$0$.

The proof of Theorem \ref{thm:matroid_spe_opt} is based on an induction on
matroids of lower
rank. After a few stages of the sequential game, we have selected a subset of
elements $X$. Notice that the resulting subgame is exactly a matroid
basis auction game in the contracted matroid $\mathcal{M}/X$. To understand such
subgames, we first prove a lemma \versions{(whose proof appears in the full version)}
that relates the \vcg prices of a player in a sequence of
contracted matroids.

\begin{lemma}
 Let $\mathcal{M}$ be a matroid, and consider a player
$i^*\in \opt(\mathcal{M})$. Consider a co-circuit $D$, and assume our auction
selects an element $k \neq i^*$, and let $\mathcal{M'}$ be the matroid that
results from
contracting $k$. Then $\vcg_{i^*}(\mathcal{M'}) \geq
\vcg_{i^*}(\mathcal{M})$ and the two are equal if $k\in \opt(\mathcal{M})$.
\label{lem:vcg_prices}
\end{lemma}
\versions{}{
\begin{proof}
First we show that the VCG prices do not change when contracting an element
from the optimum. From matroid properties it holds that for any set $X$:
$\opt(\mathcal{M}/X)\subset \opt(\mathcal{M})$. In this case
$\opt(\mathcal{M'})=\opt(\mathcal{M}/\{k\})\subset \opt(\mathcal{M})$, which
directly implies that $\opt(\mathcal{M'})=\opt(\mathcal{M})-\{k\}$.
Hence, $\{j: \opt(\mathcal{M})-i^*+j \in
\mathcal{I}_{M}\}=\{j: \opt(\mathcal{M}')-i^*+j\in
\mathcal{I}_{M}'\}$, and thus,
$\vcg_{i^*}(\mathcal{M})=\vcg_{i^*}(\mathcal{M'})$.

As mentioned in the previous section, the \vcg price can also be defined as
the maximum over all cycles of the matroid that contain $i^*$
of the minimum value bidder in each cycle: $\vcg_{i^*}(\mathcal{M}) = \max_{C\in
\mathcal{C}(\mathcal{M}): i^*\in C}\min_{i^* \neq j\in C} v_j$. Let $C$ be the
cycle
that attains this maximum in $\mathcal{M}$. The element $i^*$ is dependent on
the
set $C \setminus \{i^*\}$ in $\mathcal{M}$, and as a result $i^*$ is dependent
of the set $C \setminus \{i^*,k\}$ in $\mathcal{M'}$, hence there is a cycle
$C'\subset C$ in $\mathcal{M'}$ with $i^*\in C'$. This proves that the \vcg
price can only increase due to contracting an element.
\end{proof}
}

\versions{Next we sketch the proof of the theorem, see the full version for the details.}{Now we are ready to prove Theorem \ref{thm:matroid_spe_opt}.}

\versions{\begin{proofof}{Theorem
\ref{thm:matroid_spe_opt} (sketch)}}{\begin{proofof}{Theorem
\ref{thm:matroid_spe_opt}}}
For clarity, we assume the values of the players for being allocated to be all different.
We will prove the theorem by induction on the rank of the matroid.
Let $\mathcal{M}$ be our initial matroid prior to some auction. Notice that
for any outcome of the current auction the corresponding subgame
is exactly a sequential matroid auction on a contracted
matroid $\mathcal{M'}$.
\versions{}{
The proposed auction for rank 1 matroids is exactly a standard (no external payoffs) first price auction.

}Let $D$ be the co-circuit auctioned.
Using the induction hypothesis, we can write the induced game on this node of the game tree exactly.
For $i\in D-\opt(\mathcal{M})$, if he doesn't win the current auction then by the induction hypothesis, he is not going to win in any of the subsequent auctions, and hence $v_i^i = v_i$ and $v_i^j = 0$ for $j \neq i$.
For player a player $i \in \opt(\mathcal{M}) \cap D$. Again $v_i^i = v_i$ and for $j \neq i$, we have that $v_i^j = v_i - \vcg_i(\mathcal{M} / j)$ if $i \in \opt(\mathcal{M} / j)$ and $v_i^j = 0$ otherwise.

We claim that in all equilibria of this game, where no player bids $b_i > \gamma_i := \max_j v_i^i - v_i^j$ (notice that bidding above $\max_j v_i^i - v_i^j$ is dominated strategy), a player $i \in \opt(\mathcal{M})$ wins by his $\vcg$ price. Suppose some player $k \notin \opt(\mathcal{M})$ wins the auction. Then, there is some player $j \in  \opt(\mathcal{M}) \setminus \opt(\mathcal{M}/k)$. This is not an equilibrium, as player has $v_j > v_k$, and hence $j$ could  overbid $k$ and get the item, since the price is $p \leq v_k < v_j$.

\versions{To establish the induction hypothesis, in the full version we will show that}{Next we claim that} the winner $i \in \opt(\mathcal{M})$ gets the item by his $\vcg$ price, and the winner is $i \in D$ with highest $\vcg$ price.
\versions{}{
Suppose he gets the item by some value strictly smaller than his $\vcg$ price. If we can show that there is some player $t$ such that $v_t^t - \vcg_i(\mathcal{M}) \geq v_t^i$. Let $C$ and $j$ be the cycle and element $j$ that define the $\vcg$ price of $i$ in:
$$\vcg_{i}(\mathcal{M}) = \max_{C\in \mathcal{C}(\mathcal{M}): i\in C}\min_{i \neq j\in C} v_j$$
Now, since $\abs{C \cap D} \geq 2$, there is some $t \neq i$, $t \in C \cap D$. Notice that $v_t \geq \vcg_i(\mathcal{M})$, so if $t \notin \opt(\mathcal{M})$, then he would overbid $i$ and get the item. If $t \in \opt(\mathcal{M})$, then notice $\vcg_t(\mathcal{M}) \geq \vcg_i(\mathcal{M})$, so again he would prefer to overbid $i$ and get the item. This also shows that some player $i$ whose $\vcg$ price is not maximum winning by at most his $\vcg$ price is not possible in equilibrium.

At last, suppose the winner gets the item for some price $p$ above his $\vcg$ price. Then $b_i = p+$ and there is some player $j \in D$ such that $b_j = p$. It can't be that $j \notin \opt(\mathcal{M})$, then his value $v_j$ can't be higher than then maximum $\vcg$ price.
So, it must be that $j \in \opt(\mathcal{M})$, then player $i$ can improve his utility by decreasing his bid, letting $j$ win and win for $\vcg_i(\mathcal{M}/j) =\vcg_i(\mathcal{M}) < p$ (by Lemma \ref{lem:vcg_prices}).
}
\end{proofof}

The above optimality result tells us that $\vcg$ can be implemented in a
distributed and asynchronous way. Although the auctions happen locally, the
final price of each auction (the $\vcg$ price) is a global property. It should
be noted, nevertheless, that this is a common feature in network games in
general. The previous theorem concerns with the state of the game after
equilibrium is reached. If one considers a certain (local) dynamics and believes
it will eventually settle in an equilibrium, the $\vcg$ outcome is the only
possible such stable state.

\subsection{Unit-demand matroid auction}

In this section we sketch a common generalization of the auction for unit demand
bidders from section \ref{subsec:multi_unit_auctions} and the matroid auction of
section \ref{subsec:sequential_matroid}.
Suppose that the items
considered form the ground set of some matroid
$\mathcal{M}=([m],\mathcal{I}_\mathcal{M})$ and the auctioneer wants to sell an
independent set of this matroid, while buyers remain unit-demand and are only
interested in buying a single item.

We define the Sequential Matroid Auction with Unit-Demand Bidders to be the
game induced if in the above setting we run the Sequential First Price Auction
on co-circuits of the matroid as defined in previous section.
\comment{
We consider the following sequential
auction: at each moment pick a co-circuit that doesn't intersect any of the
previously allocated items and run a first price auction among all the players
connected to some of these items. The winner pays his bid and is
allowed to choose any element in that co-circuit that he is connected to. We
refer to such a game as a Sequential Matroid Auction with Unit-Demand Bidders.
}
%

\begin{theorem}
\label{thm:matroid-unit-demand}
The price of anarchy of a subgame
perfect equilibrium of any Sequential Matroid Auction with Unit-Demand
Bidders is $2$.
\end{theorem}


To adopt our proof from the auction with unit-demand bidder to the more general
Theorem \ref{thm:matroid-unit-demand} we define the notion of the
\textbf{participation graph} $\mathcal{P}(B)$ of a base $B$ to be a bipartite
graph between the nodes in the base and the auctions that took place. An edge
exists between an element of the base and an auction if that element
participated in the auction. Now the proof is a combination of the proof of
Theorem \ref{thm:unit-demand} and of the following lemma.

\begin{lemma} For any base $B$ of the matroid $\mathcal{M}$, $\mathcal{P}(B)$
contains a perfect matching.
\label{lem:exists_perf_match}
\end{lemma}

\versions{}{
\begin{proof}
We will prove that given any $k$-element independent set, there were $k$
auctions
that had at least one of those elements participating. Then by applying Hall's
theorem we get the lemma.

 Let $I_k=\{x_1,\ldots,x_k\}$ be such an independent set of the matroid. Let
$\mathcal{A}_{-k}=\{A_1,\ldots,A_t\}$ be the set of auctions (co-circuits) that
contain no element of $I_k$ ordered in the way they took place in the game and
$\mathcal{A}_{k}$ its complement. Let $a_1,\ldots,a_t$ be the winners of the
auctions in $A_{-k}$. Let $r(\mathcal{M})$ be the rank of the
matroid.

Since $I_k$ is an independent set, it is a subset of some basis and by the
properties of co-circuits: for any $x_i\in I_k$ there exists a co-circuit $X_i$
that contains $x_i$ and no other $x_j$. The sequence of elements
$(x_1,\ldots,x_k,a_1,\ldots,a_t)$ and co-circuits
$(X_1,\ldots,X_k,A_1,\ldots,A_t)$ have the property that each element belongs to
its corresponding co-circuit and no co-circuit contains any previous element.
Hence, the set $\{x_1,\ldots,x_k,a_1,\ldots,a_t\}$ is an independent set and
therefore
$t+k\leq r(\mathcal{M})$.
Since the total number of auctions is $r(\mathcal{M})$,
$|\mathcal{A}_{k}|\geq k$.
\end{proof}
}

\versions{}{Now the proof of Theorem \ref{thm:matroid-unit-demand}.}

\begin{proofof}{Theorem \ref{thm:matroid-unit-demand} (sketch)}
Using the last lemma, there is a bijection between the elements
 allocated in the efficient outcome and the co-circuits auctioned. For a player
 $i$ that is assigned an item $j^*(i)$ in the efficient outcome, let $A(i)$
 be the auction (co-circuit) matched with $j^*(i)$ in the above
bijection. Now if in the proof of Theorem
\ref{thm:unit-demand} we replace any reasoning about the auction of item
$j^*(i)$ with the auction $A(i)$, we can extend the arguments and
prove that $p(A(i)) \geq v_{i,j^*(i)} - v_{i,j(i)}$, where $p(A(i))$ is the value of
 the bid that won auction $A(i)$. Summing these inequalities over the auctions
completes the proof.
\end{proofof}

%


\bibliographystyle{abbrv}
\bibliography{sigproc}

\begin{appendix}

\section{Iterated elimination of dominated
strategies}\label{appendix:refinements}

First we define precisely the concept of a strategy profile that survives
iterated elimination of weakly dominated strategies. Then we characterize such
profiles for the first-price auction with externalities using a
graph-theoretical argument.

\begin{Definition}
 Given an $n$-player game define by strategy sets $S_1, \hdots, S_n$ and
utilities $u_i: S_1 \times \hdots \times S_n \rightarrow \R$ we define a
\emph{valid procedure for eliminating weakly-dominated strategies} as a
sequence $\{ S_i^t \}$ such that for each $t$ there is $i$ such that $S_j^t =
S_j^{t-1}$ for $j \neq i$, $S_i^t \subseteq S_i^{t-1}$ and for all $s_i \in
S_i^{t-1} \setminus S_i^{t}$ there is $s'_i \in S_i^t$ such that
$u_i(s'_i, s_{-i}) \geq u_i(s_i, s_{-i})$ for all $s_{-i} \in \prod_{j \neq i}
S_j^t$ and the inequality is strict for at least one $s_{-i}$. We say that an
strategy
profile $s$ survives iterated elimination of weakly-dominated strategies if for
any valid procedure $\{ S_i^t \}$, $s_i \in \cap_t S_i^t$. 
\end{Definition}

The concept above is very strong as different elimination
procedures can lead to elimination of different strategies.
This can possibly lead to no strategy (at least no Nash
equilibrium) surviving iterated elimination of weakly-dominated strategies.
We show that the first price auction game has equilibria that
satisfies this strong definition, which makes the
equilibria a very robust prediction.

As a warm up, consider the first price auction without externalities, i.e.,
$v_i^i = v_i$ and $v_i^j = 0$ for $j \neq i$ with $v_1 \geq v_2 \geq \hdots$.
It is easy to see that the set of
strategies surviving any iterated elimination procedure is $[0, v_i)$ for
player $i>1$ and $[0,v_2]$ for player $1$. Bidding $b_i > v_i$ is clearly
dominated by bidding $v_i$. By the definition, bidding $v_i$ is dominated by
bidding any value smaller then $v_i$, since by bidding $v_i$, the player can
never get positive utility. After we eliminate $b_i \geq v_i$ for all the
players, it is easy to see that $b_1 = v_2$ dominates any bid $b_1 > v_2$,
since player $1$ wins anyway (since all the other players have eliminated their
strategies $b_i \geq v_i$). The natural equilibrium to expect in this
case is player $1$ getting the item
for price $v_2$, which is a result of $b_1 = v_2+$
and $b_2 = v_2$. However, $b_2 = v_2$ is eliminated for player $2$, but any
strategy arbitrarly close to $b_2 = v_2$ is not.

This motivates us to pass to the topological closure when discussing iterative
elimination of weakly dominated strategies for first price auctions:

\begin{Definition}
In a first-price auction with externalities, 
a bid $b_i$ for player $i$ is \emph{compatible with iterated elimination of
weakly dominated strategies}, if $b_i$ is in the topological closure of the set
of bids that survive any procedure of elimination. In other words, for each
$\delta > 0$ there is a bid $b'_i$ that survives any procedure of elimination
such that $\abs{b_i - b'_i} < \delta$.
\end{Definition}

Now, we are ready to characterize the set of Nash equilibria that are
compatible with iterated elimination. In order to do that, we define an
\emph{overbidding-graph} in the following way: for each price $p$, consider a
directed graph $G_p$ on $n$ nodes
such that there is an edge from $i$ to $j$ if $v_j^j - p > v_j^i$, i.e., if
player $i$ were getting the item at price $p$, player $j$ would rather overbid
him and take the item. Now, notice that the graph $G_{p+\epsilon}$ is a
subgraph of $G_p$. 

Let's assume that all nodes have positive in-degree and
out-degree in $G_0$. If there are nodes with zero in-degree, simply remove the
players that have in-degree zero in $G_0$ (which mean that they can't possibly
want the item, i.e. they bidding zero is a dominant strategy). If there are
players with zero out-degree, then the problem is trivial, since there are nodes
for who we can give the item and get an equilibrium with zero price.

\begin{theorem}
The strategies for player $i$
that survive iterated elimination of weakly dominated strategies are $S_i = [0,
\tau_i)$ where $\tau_i$ can be computed by the following algorithm: begin with
$p=0$ and $V=[n]$. In each step, if there is a node $i \in V$ of in-degree zero
in $G_p[V]$ (i.e., $G_p$ defined on the nodes $V$), then set $\tau_i = p$ and
remove $i$ from $V$ and recurse. If there is no such node, increase the value of
$p$ until some node's in-degree becomes zero.
\end{theorem}

\begin{proof}
Consider that the players are numbered such that $\tau_1 \leq \tau_2 \leq
\hdots \leq \tau_n$. Now, we will prove by induction that no element of $[0,
\tau_i)$ can be eliminated from the strategy set of player $ j \geq i$ by
recursive elimination of weakly dominated strategies. And that there is one
procedure that eliminates all bids $b \geq \tau_i$ for player $i$ strategy set.

For the base case, suppose there is some process of iterated elimination that
removes some strategy $b \in [0, \tau_1)$ for player $i$, imagine the first time
it happens in this process and say that the strategy that eliminates it is some
$b'$. If $b' < b$, consider the profile for the other players where everyone
plays some value between $b'$ and $b$, and given that player $i$ has
positive in-degree in $G_{b}$, suppose that the highest bid is
submitted by a player $j$ such that $(j,i)$ is an edge of $G_{b}$.
Then clearly $b$ generates strictly higher utility then $b'$. Now, suppose $b' >
b$, then $b$ performs strictly better then $b'$ in the profile where all the
other players bid zero. Now, notice that all the bids $b_1 > \tau_1$ for player
$1$ are dominated and bidding $b_1 = \tau_1$ is dominated by playing any
smaller bid.

Now the induction step is along the same lines: We know that no elimination
procedure can eliminate bids in $[0,\tau_k)$ for player $k$, $k<i$. Now,
suppose there is some procedure in which we are able to eliminate some bid $b
\in [0, \tau_i )$ for some player $j \geq i$. Then again, consider the first
time it happens and let $b'$ be the bid that dominates $b$. We analyze again two
cases. If $b' < b$. consider a profile where the other players $j' \geq i, j'
\neq j$ bid between $b'$ and $b$ where the highest bidder is a player $k$ such
that the edge $(k,j')$ is in $G_b$. It is easy to see that $b$ outperforms $b'$
for this profile. If $b' > b$, we can use the same argument as in the base case.
Also, given that the strategies $b_j \geq \tau_j$ were already eliminated for
players $j < i$, clearly $b_i > \tau_i$ is dominated by $\tau_i$.
\end{proof}

\begin{corollary}
 The bids $b_i \in [0, \tau_i]$ are exactly the bids that are compatible with
iterative elimination of weakly dominated strategies for the first price
auction with externalities.
\end{corollary}

Now, given the result above, it is simple to prove that there are Nash
equilibria that are compatible with iterated elimination. Consider the
algorithm used to calculate $\tau_i$. Consider that at point $p, V$, the
active edges are the edges in $G_p[V]$. Now, in the execution of the algorithm,
we can keep track of the in-degree and out-degree of each node with respect to
active egdes. Those naturally decrease with the execution of the algorithm.
Since in each step some edges become inactive, there is at least one node such
that its out-degree becomes zero before or at the same time that his in-degree
becomes zero. So, for the corresponding player $i$, there is one price $p$ such
that $\tau_i \geq p$, there is an edge $(j,i)$ in $G_{p'}[V']$, where $p',V'$
is the state of the algorithm just before the out-degree of $i$ became zero.
So, clearly $\tau_j \geq p$. Now, it is easy to see that the strategy profile
$b_i = p+$, $b_j = p$ and $b_k = 0$ for all $k \neq i,j$ is a Nash equilibrium
and it is compatible with iterative elimination.

In fact, the reasoning above allows us to fully characterize and enumerate all
outcomes that are a Nash equilibrium compatible with iterated elimination:

\begin{theorem}
The outcome of player $i$ winning the item for price $p$ can be expressed as a
Nash equilibrium that is compatible with iterated elimination iff $p \leq
\tau_i$, player $i$ has out-degree zero in $G_p$ and there is some player $j$
with $\tau_j \geq p$ such that the edge $(j,i)$ is in $G_{p'}$ for all $p' < p$.
\end{theorem}

\section{Formal definition of extensive form games}\label{appendix:extensive}

We provide in this session a formal mathematical description of the concepts
described in session \ref{sequential-item-auctions}: We can represent an
\textbf{extensive-form game} via a game-tree, where nodes of
the tree correspond to different histories of play. At each stage of the
game, players make simultaneous moves, that can depend on the history of play so
far. So a player's strategy in an extensive form game is a strategy for each
possible history, i.e., each node of the tree. More formally, 
  
\begin{itemize}
 \item Let $N$ denote the set of players, and let $n=\abs{N}$
 \item A $k$-stage game is represented by a directed game tree $\mathcal{T} =
(V,E)$ of $k+1$ levels. Let $V^t$ be the nodes in level $t$, where $V^t$ denotes
possible partial histories at the start of stage $t$. So $V^1$ contains only
the root and $V^{k+1}$ contains all the leaves, i.e., the outcomes of the game.
Note that the tree can be infinite, if for example some player has an infinite
strategy set. 
 \item for each $v\in V \setminus V^{k+1}, i \in N$, a strategy set $S_i(v)$ is 
the set of possible strategies of player $i$
 \item for each $v \in V$, the out-going edges of $v$ correspond to strategy
profiles $s(v) \in \times_i S_i(v)$, the outcome of this stage when players
play strategies $s(v)=(s_1(v),\ldots,s_n(v))$. 
 \item for each $i \in N$, we have the utility function $u_i : V^{k+1} 
\rightarrow \R$, that denotes the utility of the outcome corresponding to node
$v \in V^{k+1}$ for player $i$.
\end{itemize}

The \textbf{pure strategy} of a player consists of choosing $s_i(v) \in S_i(v)$
for each node $v \in V$, i.e. a function $s_i:V \rightarrow \cup_v S_i(v)$ such
that $s_i(v) \in S_i(v)$. In other words, it is a strategy choice for each
round, given the history of play so far, which is encoded by node $v$. A
strategy profile is a $n$-tuple $s = (s_1, \hdots, s_n)$. It defines the
\textbf{actual history of play } $h = (h_1, h_2, \hdots, h_{k})$, where $h_1 =
s(r)$ is the strategy profile played at the root, and $h_i$ is the strategy
profile played at the node that corresponds to history $h_1, \hdots,
h_{i-1}$. Notice that $h$ corresponds to a leaf of the tree, which allows to
define the utility of $i$ for a strategy profile:
$$u_i(s) = u_i(h(s))$$

\comment{
A \textbf{mixed strategy} corresponds to choosing $\sigma_i(v) \in
\Delta(S_i(v))$ for each $v \in V$, where $\Delta(X)$ is the set of
distributions over the elements of $X$ \footnote{One could alternatively define
a mixed strategy as a distribution over pure strategies, but those definitions
are equivalent, according to Kuhn's Theorem \cite{kuhn53}.}. A mixed strategy
profile is defined by $\sigma = (\sigma_1, \hdots, \sigma_n)$. Now, one can
define history of play in the same way: the difference now is that $h$ is a
random variable. We define:
$$u_i(\sigma) = \E u_i(h(\sigma))$$
}

We use \textbf{subgame perfect equilibrium} ($\spe$) as our main solution
concept.
A subgame of sequential game is the game resulting after fixing some initial
history of play, i.e., starting the game from a node $v$ of the game tree. 
Let $u^v_i(s)$ denote the utility that $i$ gets from playing $s$
starting from node $v$ in the tree.
We say that a profile $s$ is a \spe if it is a Nash equilibrium for each subgame
of the game, that is, for all nodes $v$ we have:
$$\forall s'_i: u_i^v(s_i, s_{-i}) \geq u_i^v(s'_i, s_{-i}).$$

Given a node $v$ in the game tree and fixing $s_i(v')$ for all $v'$ below $v$, 
we can define an induced normal-form game in node $v$ by $s$ as the game with
strategy space $\times S_i(v)$ such that the utility for player $i$ by playing
$\tilde{s}(v), \tilde{s}_i(v) \in S_i(v)$ is $u_i^v(s_i, s_{-i})$ where player
$i$ plays $\tilde{s}_i(v)$ in node $v$ and according to $s_i(v')$ in all nodes
$v'$ below $v$. Kuhn's Theorem states that $s$ is a subgame perfect equilibrium
iff $s(v)$ is a Nash equilibrium on the induced normal-form game in node $v$ for
all $v$.

\comment{
We say that an $\spe$ is an \textbf{undominated subgame perfect equilibrium} if 
the equilibrium selected in the induced game in each node of the tree is in
undominated strategies. I.e., conditioned on profiles selected in nodes below
$v$, no player $i$ employs a strategy $s_i(v)$  such that there is $s'_i(v)$ in
node $v$ such that for all possible choices of $s_{-i}(v)$, $s'_i(v)$ generates
at least as much utility as $s_i(v)$ and outperforms it for some $s'_i(v)$.
Notice this definition is equivalent to apply a single round of elimination of
weakly-dominated strategies in the induced subgame. An \textbf{iteratively
undominated subgame perfect equilibrium} is one that in each induced node-game
the equilibrium survives any order of elimination of dominated strategies.
}

The main tool we will use to analyse those games is the \textbf{price of
anarchy}. Consider a welfare function defined on the leaves of the tree, i.e. $W
: V^{k+1} \rightarrow \R$. Given a certain strategy profile $s$ and its induced
history $h(s)$, the social welfare of this game play is given by $W(v) = \sum_{i
\in N} u_i(h(s))$. We define the optimal welfare as $W^* = \max_{v \in V^{k+1}}
W(v)$, and the pure Price of Anarchy ($\poa$) as:
$$\poa = \max_{s \in E} \frac{W^*}{W(s)}$$
where $E$ is the set of all subgame perfect equilibria.

There sequential auctions we study are $m$-stage games and strategy space on 
each node $v$ for player $i$ is a bid $b_i(v) \in [0, \infty)$. In other words,
the strategy of each player in this game is a function that maps the bid
profiles in the first $k-1$ items to his bid in the $k$-th item. Their utility
is the total value they get for the bundle they acquired minus the price paid.
The welfare is the sum of the values of all players.

\section{Non-Existense of SPE in Multi-Item Auctions}\label{appendix:multi-unit}

We give an example of a multi-item sequential auction with no $\spe$ in pure
strategies. The example has $4$ players and $5$ items. The first two items
$X_1,X_2$ are auctioned simultaneously first and the remaining items are
auctioned sequentialy afterwards in the order $W,Y,Z$. Players $1$ and $4$ are
single minded. Player $1$ has value $v$ only for item $Z$ and player $4$ has
value $\frac{2\delta}{3}+\epsilon$ only for item $W$. Players $2$ and $3$ have
coverage submodular valuations that are depicted in Figure \ref{fig_non_vals}.
One can check that the following allocation and prices constitutes a walrasian
equilibrium of the above instance: $A_1=\emptyset, A_2 = \{X_1,Z\},
A_3=\{X_2,Y\}, A_4=\{W\}), p_{X_1}=p_{X_2}=\delta/3,
p_{Y}=v+\delta/6,p_{W}=2\delta/3$. However, we will show that there is no
subgame perfect equilibrium in pure strategies. 

\begin{figure}
\centering
\includegraphics{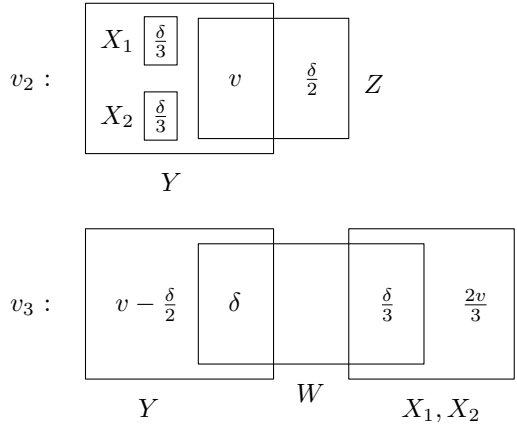}
\caption{Valuations $v_2$ and $v_3$.}
\label{fig_non_vals}
\end{figure}

We will show that the subgame perfect equilibrium in the last three auctions 
is always unique given the outcome in the first two item auction and is such
that player $1$ has a huge value for winning both $X_1,X_2$ and almost $0$
otherwise and player $3$ has huge value for winning any of $X_1$ or $X_2$ and
almost $0$ otherwise. Thus ignoring players $2$ and $4$ since they have
negligent value for $X_1,X_2$ we observe that the first two-item auction is an
example of an AND and an OR bidder that is well known to not have walrasian
equilibria and hence pure nash equilibria in the first price item auction.

So we examine what happens after any outcome of the first two-item auction:
\begin{itemize}
 \item Case 1: Player $1$ won both $X_1,X_2$.

In this case player $2$ has a value of $v+\frac{2\delta}{3}$ for $Y$ and a value
of $v+\delta/2$ for $Z$ given that he loses $Y$. In the $Z$ auction player $2$
will bid $v$. Hence, player $2$ will gain a profit of $\delta/2$ from the $Z$
auction if he loses $Y$. Moreover, the value of player $3$ for $W$ is
$\delta+\delta/3$.
    \begin{itemize}
      \item Case 1a: Player $3$ won $W$. 
      In this case the value of $3$ for $Y$ is $v-\delta/2$. Hence, the game
played at the $Y$ auction is the following (we ignore player $4$):
$$[v_i^j] = 
 \begin{bmatrix} 0 & v-\frac{\delta}{2} & 0\\ 
                 \delta/2 & v+\frac{2\delta}{3} & \delta/2\\ 
		 0 & 0 & v-\delta/2 \end{bmatrix} $$
      Thus player $2$ wants to win for a price of at most
$v+\frac{2\delta}{3}-\frac{\delta}{2}$. Player $3$ will bid $v-\delta/2$ and
player $2$ will win. In the last auction player $2$ will just bid $\delta/2$.
Hence, player $1$ will get utility $v-\delta/2$, player $2$ utility
$\frac{\delta}{2}+\frac{2\delta}{3}$ and player $3$ utility $0$.
      \item Case 1b: Player $3$ lost $W$.
      In this case the value of $3$ for $Y$ is $v+\delta/2$ and the
game played is:
$$[v_i^j] = 
 \begin{bmatrix} 0 & v-\frac{\delta}{2} & 0\\ 
                 \delta/2 & v+\frac{2\delta}{3} & \delta/2\\ 
		 0 & 0 & v+\delta/2 \end{bmatrix} $$
     Thus player $3$ now wants to win for a value at most
$v+\delta/2$ and player $2$ for a value at most
$v+\frac{2\delta}{3}-\frac{\delta}{2}$. Hence, in the unique no-overbidding
equilibrium player $3$ will win. Therefore, player $1$ will get utility
$0$,player $2$ utility $\delta/2$ and player $3$ utility $\frac{\delta}{3}$.
     \end{itemize}
 Thus we see that at auction $W$ the following game  is played:
$$[v_i^j] = 
 \begin{bmatrix} 0 & 0 & v-\frac{\delta}{2} & 0\\ 
                 \frac{\delta}{2} & \frac{\delta}{2} &
\frac{\delta}{2}+\frac{2\delta}{3} & \frac{\delta}{2}\\ 
		 \frac{\delta}{3} & \frac{\delta}{3} & \delta+\frac{\delta}{3} &
\frac{\delta}{3} \\
		 0 & 0 & 0 & \frac{2\delta}{3}+\epsilon \end{bmatrix} $$
  Players $1$ and $2$ will bid $0$ and player $4$ wants to win for at most
$\frac{2\delta}{3}+\epsilon$. Player $3$ wants to win for at most $\delta$.
Hence, in the unique equilibrium player $3$ will win $W$. Consequently, player
$2$ will win $Y$ and player $1$ will win $Z$. Thus, player $1$ will get utility
$v-\frac{\delta}{2}$, player $2$ utility $\frac{2\delta}{3}$, player $3$
utility $\frac{2\delta}{3}-\epsilon$ and player $4$ utility $0$.

\item Case 2: Player $3$ won at least one of $X_1$ or $X_2$.
    
In this case player $2$ has a value of at least $v+\frac{\delta}{3}$ and at
most $v+\frac{2\delta}{3}$ for $Y$ and a value of $v+\delta/2$ for $Z$ given
that he loses $Y$. In the $Z$ auction player $2$ will bid $v$. Hence, player $2$
will gain a profit of $\delta/2$ from the $Z$ auction if he loses $Y$. Moreover,
the value of player $3$ for $W$ is
$\delta$.
    \begin{itemize}
      \item Case 2a: Player $3$ won $W$. 
      In this case the value of $3$ for $Y$ is $v-\delta/2$. Hence, the game
played at the $Y$ auction is the following (we ignore player $4$):
$$[v_i^j] = 
 \begin{bmatrix} 0 & v-\frac{\delta}{2} & 0\\ 
                 \delta/2 & v+\frac{\delta}{3} \text{~or~} v+\frac{2\delta}{3} &
\delta/2\\ 
		 0 & 0 & v-\delta/2 \end{bmatrix} $$
      Thus player $2$ wants to win for a price of at most
$v+\frac{\delta}{3}-\frac{\delta}{2}$. Player $3$ will bid $v-\delta/2$ and
player $2$ will win. In the last auction player $2$ will just bid $\delta/2$.
Hence, player $1$ will get utility $v-\delta/2$, player $2$ utility
$\geq \frac{\delta}{2}+\frac{\delta}{3}$ and player $3$ utility $0$.
      \item Case 2b: Player $3$ lost $W$.
      In this case the value of $3$ for $Y$ is $v+\delta/2$ and the
game played is:
$$[v_i^j] = 
 \begin{bmatrix} 0 & v-\frac{\delta}{2} & 0\\ 
                 \delta/2 & v+\frac{\delta}{3} \text{~or~} v+\frac{2\delta}{3} &
\delta/2\\ 
		 0 & 0 & v+\delta/2 \end{bmatrix} $$
     Thus player $3$ now wants to win for a value at most
$v+\delta/2$ and player $2$ for a value at most
$v+\frac{2\delta}{3}-\frac{\delta}{2}$. Hence, in the unique no-overbidding
equilibrium player $3$ will win. Therefore, player $1$ will get utility
$0$, player $2$ utility $\delta/2$ and player $3$ utility at least
$\frac{\delta}{3}$.
     \end{itemize}
 Thus we see that at auction $W$ the following game is played:
$$ \begin{bmatrix} 0 & 0 & v-\frac{\delta}{2} & 0\\ 
                 \frac{\delta}{2} & \frac{\delta}{2} &
\frac{\delta}{2}+\frac{\delta}{3} \text{~or~}
\frac{\delta}{2}+\frac{2\delta}{3} & \frac{\delta}{2}\\ 
		 \frac{\delta}{3} \text{~or~} \frac{2\delta}{3} &
\frac{\delta}{3} \text{~or~} \frac{2\delta}{3} &
\delta &
\frac{\delta}{3} \text{~or~} \frac{2\delta}{3} \\
		 0 & 0 & 0 & \frac{2\delta}{3}+\epsilon \end{bmatrix} $$
  Players $1$ and $2$ will bid $0$ and player $4$ wants to win for at most
$\frac{2\delta}{3}+\epsilon$. Player $3$ wants to win for at most
$\delta-\frac{\delta}{3}$.
Hence, in the unique equilibrium player $4$ will win $W$. Consequently,
player $3$ will win $Y$ and player $2$ will win $Z$. Thus, player $1$ will get
utility $0$, player $2$ utility $\delta/2$, player $3$
utility at least $\frac{\delta}{3}$ and at most $\frac{2\delta}{3}$ and player
$4$ utility at least $\epsilon$ and at most $\frac{\delta}{3}+\epsilon$
(according to whether player $2$ won one of $X_1,X_2$ or not).

\item Case 3: Player $3$ didn't win any of $X_1,X_2$ and player $2$ won some of
$X_1,X_2$.

In this case we just need to observe that $2$ expects a profit of at most
$\delta/2$ from $Z$ hence he will set a price of at least $v-\delta/2$ at the
$Y$ auction. Thus player $3$ expects to get utility at most $2\delta$ from the
$Y$ and $W$ auctions. 
\end{itemize}

Now we examine the existence of equilibrium in the two-item auction. Both
players $2$ and $4$ get utilities at most $2\delta$ from the $Y,W$ and $Z$
auctions and have at most $\delta$ value for $X_1$ and $X_2$. Thus they will
bid at most $3\delta$.
On the other hand player $1$ has a utility of $v-\delta/2$ from subsequent
auctions if he wins both items and utility $0$ if player $3$ wins some of them.
Moreover, player $3$ has a utility at most $2\delta$ from subsequent auctions
in any outcome, but has a value of $2v/3+\frac{\delta}{3}$ for winning some of
$X_1$ or $X_2$. Hence, player $3$ is willing to win some of $X_1$ or $X_2$ at a
price of $2v/3-2\delta$. Since we assume that $\delta\rightarrow 0$ we can
ignore players $2$ and $4$ in the first auction.

If player $1$ wins both items and both at a price smaller than
$2v/3-2\delta$ then player $3$ has a profitable deviation to bid higher than
that at one auction and outbid $1$. Thus if player $1$ wins both items he must
be paying at least $4v/3-4\delta$ which is much more than the utility he
receives. Hence, this cannot happen.

Thus player $3$ must be winning some auction. If that is true then player $2$
receives $0$ utility in any possible outcome and since he has no direct value
for $X_1$ or $X_2$ he doesn't won to win any of the auctions. Moreover, if
player $3$ bids more than $2\delta$ in both auctions and wins both auctions then
he has a profitable deviation to bid $0$ in one of them since given that he wins
one item his marginal valuation for the second is $0$. Thus in equilibrium
player $3$ will bid less than $2\delta$ in some of the two auctions. Moreover,
he is bidding at most $2v/3+2\delta$ in the auction he is winning. However, in
that case player $1$ has a profitable deviation of marginally outbidding player
$3$ in both auctions. Hence, player $3$ winning some auction cannot happen
either at equilibrium and therefore no pure
nash equilibrium can exist in the first round.

\section{Second vs First price in Sequential Auctions}
\label{sec:second-price}

In order to stress how essential is the design decision of adopting first price instead of second price in the sequential auctions\footnote{or altenatively, how crucial the envy-free assumption in second-price auctions is}, we present two examples that show how sequential second price auction fail to provide any welfare guarantee even for elementary valuations. It is important to notice that this happens even though we restrict ourselves to equilibria where no player overbids in any game induced in a node of the game-tree. The second example is even stronger: even if we restrict out attention to equilibria that remain after iterative elimination of weakly dominated strategies of all induced games, still no welfare guarantee is possible.

\subsection{Additive Valuations}

Consider a sequential auction of $m$ items among $n$ players using a sequential second price auction, where each player has additive valuation $v_i:2^{[m]} \rightarrow \R_+$, i.e., $v_i(S) = \sum_{j \in S} v_j(\{j\})$. It is tempting to believe that this is equivalent to $m$ independent Vickrey auctions. Using $\spe$ as a solution concept, however, allows the possibility of signaling. 

Consider the following example with $3$ players, where the Price of Anarchy is
infinite, which happens due to a miscoordination of the players. Consider $t+2$
items $\{A_1, \hdots, A_t, B, C\}$ and valuations given by the following table:

\begin{center}
\begin{tabular}{ c || c | c | c | c | c | c  }
    & $A_1$ & $A_2$ & $\hdots$ & $A_t$ & $B$ & $C$ \\
  \hline                       
  $1$ & $1$ & $1$ & $\hdots$ & $1$ & $0$ & $1$ \\
  $2$ & $1-\epsilon$ & $1-\epsilon$ & $\hdots$ & $1-\epsilon$ & $1$ & $1-\epsilon$ \\
  $3$ & $\delta$ & $\delta$ & $\hdots$ & $\delta$ & $1-\epsilon$ & $0$ \\
\end{tabular}
\end{center}

Now, notice that in each subtree, it is an equilibrium if everyone plays truthfully in the entire subtree and notice that under this players get only very small utility. Now, consider the outcome where player $3$ gets items $A_1 \hdots A_t$, player $2$ gets item $B$ and player $1$ gets item $C$. This outcome has social welfare $SW = 2 + t\epsilon$ while $Opt = t + 2$. Now we argue that there is an $\spe$ that produces this outcome, showing therefoe that the Price of Anarchy is unbounded.

In the game tree, in the path corresponding to the equilibrium described above, consider the winner bidding truthfully and all other bidding zero. Now, in all other decision nodes of the tree outside that path, let everyone bid truthfully. It is easy to check that this is a $\spe$ according to the definition above.

Notice that this is a feature of second price. For example, in the last auction, player $3$ couldn't have gotten this item for free in the first price version, since player $2$ would have been overbidded him and got it instead. Second price auctions have the bug that a player can win an item for some price $p$, but some other player to take the item, he may need to pay $p' > p$ and it may make (as in the example) the equilibrium be non-envy free.

\subsection{Unit-demand players}

In this section we present a unit-demand sequential second price instance that
exhibits arbitrarily high $\spoa$. The instance we present involves signaling
behaviour from the players. Moreover, the second price nature of the auction
enables players to signal for a zero price and as much as they want, a
combination that has devastating effects on the efficiency. 

\begin{figure}[h]
\centering
\includegraphics{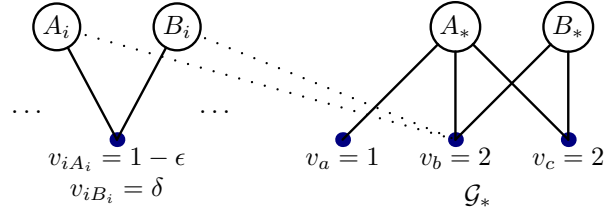}
\caption{Sequential Second Price instance with high $\spoa$. Auctions happen
from left to right. If a bidder is not connected to an item that implies $0$
value. If a bidder has a single value then that is his value for any of
the items he is connected to. Dashed lines mean $0$ value but imply that the
bidder might bid at that auction despite the $0$ value.}
\label{fig2}
\end{figure}

The instance is depicted in Fig. \ref{fig2} is an auction with $2k+2$ items
auctioned in the order $A_1, B_1, A_2, B_2, \hdots, A_k, B_k, A_*, B_*$ and
$n+3$ player called $1,2,\hdots,k,a,b,c$. The main component of the
instance
is gadget $\mathcal{G}_{*}$, which comprises of the last two auctions of the
game. As a subgame $\mathcal{G}_{*}$ has two possible subgame perfect
equilibria: In the first equilibrium, which we denote
$\spe_1$, $b$ wins $A_*$ at price $1$ and $c$ wins $B_*$ at price $0$. In the
second $\spe$ $c$ wins $A_*$ and $b$ wins $B_*$. Hence, player $b$'s utility
is $1$ unit higher in $\spe_1$. 

In what follows we construct a $\spe$ of the whole instance that survives
iterated elimination of weakly dominated strategies and exhibits unbounded
price of anarchy. We describe what happens in the last $2$ external auctions
$A_k,B_k$. If player $b$ or $c$ win at auction $A_k$ and at $0$ price then in
last two auctions $\spe_1$ is implemented. If player $k$ wins auction $A_k$ then
$\spe_2$ is implemented. If player $k$ loses and sets a positive price then
if either $b$ or $c$ win at auction $B_k$ then $\spe_1$ is implemented otherwise
$\spe_2$. Now using backwards induction we see that if player $c$ has no
incentive to bid at any of $A_k,B_k$. Moreover, if player $b$ wins $A_k$ at any
price then at $B_k$ he has a utility of $2$ for winning and $1$ for losing.
Thus at $B_k$ he bids $1$ and player $k$ bids $\delta$. Thus, at $A_k$ player
$b$ has a utility of $1-\epsilon$ for winning at any price. Hence, he will bid
$1-\delta>1-\epsilon$. Now, player $k$ knows that he is going to lose at $A_k$,
and if he sets a positive price he is going to also lose at $B_k$. On the other
hand if he sets a price of $0$ at $A_k$ then none of $b,c$ have any incentive
to outbid him on $B_k$ which will give him a utility of $\delta$. Thus, player
$k$ will bid $0$ on $A_k$. We can copy this behaviour by adding several
auctions $A_i,B_i$ happening before $A_*,B_*$. At each of these auctions player
$b$ is going to be winning auction $A_i$ at a price of $0$ and the
corresponding player $i$ will be winning auction $B_i$. This leads to a $\spoa=
\frac{k(1-\epsilon)+4}{k\delta+4}=O(\frac{1-\epsilon}{\delta})$ which can 
be arbitrarily high.

\end{appendix}


\end{document}